\documentclass[preprint,superscriptaddress,notitlepage,endfloat,titlesec,longbibliography]{revtex4-1}
\usepackage{graphicx,bm,mathtools,color}
\usepackage{lineno}

\begin{document}

\title{Anomalous quantum oscillations and evidence for a non-trivial Berry phase in SmSb}
\author{Fan Wu}
\affiliation{Center for Correlated Matter and Department of Physics, Zhejiang University, Hangzhou 310058, China}
\author{Chunyu Guo}
\affiliation{Center for Correlated Matter and Department of Physics, Zhejiang University, Hangzhou 310058, China}
\author{Michael Smidman}
\affiliation{Center for Correlated Matter and Department of Physics, Zhejiang University, Hangzhou 310058, China}
\author{Jinglei Zhang}
\affiliation{Anhui Province Key Laboratory of Condensed Matter Physics at Extreme Conditions, High Magnetic Field Laboratory of the Chinese Academy of Sciences, Hefei 230031, Anhui, China}
\author{Ye Chen}
\affiliation{Center for Correlated Matter and Department of Physics, Zhejiang University, Hangzhou 310058, China}
\author{John Singleton}
\affiliation{National High Magnetic Field Laboratory, MS-E536, Los Alamos National Laboratory, Los Alamos, NM 87545, U.S.A.}
\author{Huiqiu Yuan}
\email{hqyuan@zju.edu.cn}
\affiliation{Center for Correlated Matter and Department of Physics, Zhejiang University, Hangzhou 310058, China}
\affiliation{Collaborative Innovation Center of Advanced Microstructures,Nanjing University, Nanjing 210093, China}

\date{\today}

\begin{abstract}
	
	Topologically non-trivial electronic structures can give rise to a range of unusual physical phenomena, and the interplay of band topology with other effects such as electronic correlations and magnetism requires further exploration. The rare earth monopnictides $ X $(Sb,Bi) ($ X $ = lanthanide) are a large family of semimetals where these different effects may be tuned by the substitution of rare-earth elements. Here we observe anomalous behavior in the quantum oscillations of one member of this family, antiferromagnetic SmSb. The analysis of Shubnikov-de Haas (SdH) oscillations provides evidence for a non-zero Berry phase, indicating a non-trivial topology of the $\alpha$-band. Furthermore, striking differences are found between the temperature dependence of the amplitudes of de Haas-van Alphen effect oscillations, which are well fitted by the Lifshitz-Kosevich (LK) formula across the measured temperature range, and those from SdH measurements which show a significant disagreement with LK behavior at low temperatures. Our findings of unusual quantum oscillations in an antiferromagnetic, mixed valence semimetal with a possible non-trivial band topology can provide an opportunity for studying the interplay between topology, electronic correlations and magnetism.

\end{abstract}

\maketitle

\section*{Introduction}

The discovery of topological insulators with bulk band gaps but robust topological surface states protected by time reversal symmetry triggered much research into topological phases of matter \cite{zhang2009topological,chen2009experimental}. Topological insulators are bulk insulators yet have conducting surface states as a consequence of the topologically non-trivial band structure, where a Dirac cone lies within the bulk gap \cite{Hre,TSCTI}. Subsequently, a number of semimetals with non-trivial topologies have also been discovered. In some cases such as Dirac and Weyl semimetals, the electronic structure is gapless and contains Dirac or Weyl points at the positions where the bands cross with linear dispersions and the excitations are well described as Dirac or Weyl fermions \cite{Liu,xiong2015evidence,borisenko2014experimental,CavaCd3As2,Weyl1,Weyl2,lv2015experimental}. Meanwhile in other topological semimetals, the band structure contains similar band inversion features to topological insulators, but the Fermi level does not lie within the band gap \cite{chadov2010tunable,liu2016observation}. The role of electronic correlations in topological systems has also become of particular interest after the proposal that SmB$_6$ is a topological Kondo insulator, where the surface state lies within the gap opened due to the Kondo interaction \cite{Dzero2010,kim2013surface,wolgast2013}. One striking feature is the presence of quantum oscillations in SmB$_6$ from de Haas-van Alphen effect (dHvA) measurements \cite{Li1208,tan2015unconventional}, despite the material being a bulk insulator at low temperatures, stimulating debate over whether these arise from the surface \cite{Li1208,erten2016kondo}, or if this is a bulk property \cite{tan2015unconventional,knolle2017excitons}. Furthermore, Tan \textit{et al}. found that the dHvA amplitude shows an anomalous increase at low temperatures below 1~K, and the origin of this highly unusual deviation is currently not resolved \cite{tan2015unconventional}.

Recently, the $X$(Sb,Bi)($X$~=~lanthanide) series of semimetals with a cubic rocksalt structure have attracted considerable interest. These materials are commonly found to have an extremely large magnetoresistance (XMR), generally attributed to the nearly perfect compensation of electron and hole carriers \cite{LaSbCava,Cavatwo,LaBiPRB,he2016distinct,2016arXiv161102927G,NdSbPRB,YSbSr}. Meanwhile, non-trivial band topologies have been theoretically predicted \cite{zeng2015topological,calculation}, and evidence for topological states is found experimentally for some compounds from ARPES and transport measurements \cite{CeSbARPES,2016arXiv161102927G,YSbSr,NdSbIOP,Nayak2017Multiple,LaBiARPES,LaXFDL,XBiARPES}, while other members appear to be topologically trivial \cite{DHLaSb,he2016distinct,wu2017extremely}. In LaSb, it was suggested that the low-temperature plateau in the temperature dependence of the resistivity in applied magnetic fields was due to a topologically protected surface state, similar to that observed in SmB$ _6 $ \cite{LaSbCava}. However, ARPES measurements show conflicting results over whether the band topology is trivial \cite{DHLaSb,LaXFDL}, while the resistivity plateau can be explained by a three band model with nearly perfect electron-hole compensation \cite{DHLaSb}. Furthermore, by substituting for lanthanide elements with partially filled $4f$ electron shells, the effect of tuning the spin-orbit coupling, magnetism and electronic correlations can be studied. For example, CeSb shows evidence for Weyl fermions in the field-induced ferromagnetic state from both theoretical calculations and the measurement of a negative longitudinal magnetoresistance \cite{2016arXiv161102927G}. NdSb has also been proposed to be a topological semimetal from the observation of a Dirac-like semi-metal phase using ARPES, as well as from band structure calculations and analysis of the Landau indices \cite{NdSbIOP,NdSbHField}. 

In this work, we report magnetoresistance and quantum oscillation measurements of the antiferromagnet SmSb \cite{mullen1974}. Based on the analysis of Shubnikov-de Haas (SdH) oscillations, we found that the phase of all the harmonic oscillations corresponding to the $\alpha$-band is close to $\pi$, while for the other bands it is almost zero. These results provide evidence that the band topology of the $\alpha$-band is non-trivial. Moreover, striking differences are found between the temperature dependences of the quantum oscillation amplitudes from SdH and dHvA measurements. The amplitudes of the SdH oscillations show anomalous behavior at low temperatures, which may arise due to the presence of multiple Fermi surface sheets or a conduction channel related to the topological state.

\section*{Results and Discussion}

The temperature dependence of the electrical resistivity [$\rho(T)$] of SmSb is displayed in Fig.~1 \textbf{a}, where the data up to 300~K is shown in the inset. The $\rho(T)$ data show a small hump at around 65~K, which is explained as resulting from the splitting of the crystalline electric fields \cite{beeken1978intermediate}, as well as a sharp drop of $\rho(T)$ at the antiferromagnetic transition at $T_N=2.2$~K \cite{hulliger1978low}. While $T_N$ changes very little in applied magnetic fields \cite{ozeki1991haas}, an upturn appears in $\rho(T)$ at low temperatures, which becomes more pronounced as the field is increased. As shown in Fig.~1\textbf{a} and 1\textbf{b}, there is a strong increase of $\rho(T)$ in-field below $T_N$, before reaching a nearly constant value below around 1~K. Figure~1\textbf{c} shows the field dependence of the magnetoresistance [$\rho(H)/\rho(0)$] of SmSb up to 15~T at various temperatures down to 0.3~K. At 0.3~K $\rho(H)/\rho(0)=74000$, which decreases rapidly as the temperature is increased. Figure~1\textbf{d} displays measurements performed up to higher fields at 1.8~K, where the magnetoresistance increases quadratically to about 5558 at 60~T. The magnetoresistance of SmSb is one of the largest observed among the $X$Sb family of compounds \cite{DHLaSb,NdSbHField,2016arXiv161102927G,ye2017extreme,wu2017extremely}, and the quadratic field dependence suggests that the XMR most likely results from electron-hole carrier compensation \cite{PippardMR}. In this case, the increase of mobility below $T_N$ may lead to both the small low temperature values of $\rho(T)$ in zero-field, as well as the very large magnetoresistance and low-temperature plateau of $\rho(T)$ in high fields. While this low temperature plateau has also been observed in other $X$Sb materials \cite{LaSbCava,2016arXiv161102927G,YSbSr,NdSbPRB}, the appearance of the plateau at a lower temperature in SmSb is likely due to the comparatively low $T_N$.

To examine the topology of the electronic structure of SmSb, we also analyzed the SdH oscillations in the resistivity for the two samples displayed in Fig.~1\textbf{d}. From the fast Fourier transformation analysis (see Supplementary Figure 1), the most prominent oscillation frequencies consist of three fundamental frequencies ($F_{\alpha}$, $F_{\beta}$ and $F_{\gamma}$), and two harmonic frequencies ($F_{2\alpha}$ and $F_{3\alpha}$). In the following, the data are analyzed with five oscillations, corresponding to these frequencies. Note that for the data measured in a pulsed magnetic field, the data measured at 15~K were analyzed, so as to avoid complications arising from strong harmonic oscillations. To obtain the oscillation phase factors ($\phi^i$) corresponding to each fundamental frequency, we have used Lifshitz-Kosevich (LK) theory to fit the data \cite{PRXAlex}:
\begin{eqnarray}
\Delta\rho/\rho_0 = \sum_{i=1}^{3}\sum_{r}a_{i,r}\sqrt{B/r}R_T^{i,r}R_D^{i,r}R_S^{i,r}{\rm cos}(2\pi r(F_i/B-1/2)-\phi^i).
	\label{equation1}
\end{eqnarray}

Here $a_{i,r}$ is a constant corresponding to the $r$th harmonic of the $i$th band, $\phi^i$ is the corresponding phase factor, and $R_T$, $R_D$ and $R_S$ are the temperature, Dingle and spin damping factors respectively. These are given by $ R_T^{i,r}=(14.69rm_iT/B)/{\rm sinh}[(14.69rm_iT/B)] $, $ R_D^{i,r}={\rm exp} (-14.69rm_iT_D/B) $, and $ R_S^{i,r}={\rm cos}(\pi grm_i/m_e) $, where $m_i$ denotes the cyclotron mass for each band, $ m_e $ is the free electron mass and $ T_D $ is the Dingle temperature. The data were fitted using Eq.~\ref{equation1}, and it can be seen in Fig.~2 that this can well describe the SdH oscillations both measured in a static field up to 35~T (S2) and a pulsed field up to 60~T (S6). Note that since the data are analyzed for a fixed temperature and field-angle, $R^{i,r}_{S}$ is taken to be a constant value. Here $F_i$ for the five components were fixed to the values from the FFT analysis, and the $m_i$ for the $\alpha-$ and $\beta-$ bands were obtained from the temperature dependence of the oscillation amplitudes from dHvA measurements (see Fig.~5). On the other hand it is difficult to obtain $m_i$ from analyzing the temperature dependence of oscillations corresponding to the $\gamma-$band, and therefore this was a fitted parameter in the analysis. In addition, $T_D$ and $\phi^i$ for each band were fitted parameters in the analysis, and the results are displayed in Supplementary Table I.

The phase factor corresponding to the $\alpha$-band from both samples is 1.16$\pi$. Note that the phase is given by $\phi^i = \phi_D^i+\lambda^i$, where $\phi_D$ corresponds to the LK correction, and $\lambda$ is the sum of geometric and dynamical phases, which includes the Berry phase, the phase factor resulting from the orbital magnetic moment and the Zeeman coupling phase factor \cite{PRXAlex}. For a two-dimensional cylinder-like Fermi surface, $\phi_D$ is zero, while for a three-dimensional Fermi surface pocket, $\phi_D$ is $\pi/4$ or $-\pi/4$ for oscillations corresponding to the minimum or maximum Fermi surface cross sections respectively. Since the oscillation frequency is lowest when the field is applied along the [100] direction (see Supplementary Figure 2), this suggests that $\phi_D=\pi/4$, and hence $\lambda=0.91\pi$. This is close to the value of $\pi$ indicating a $\pi$ Berry phase, as expected for a topologically non-trivial electronic band \cite{mikitik1999manifestation,PhysRevLett93166402,shoenberg2009magnetic}, while the small deviation from $\pi$ suggests the possible influence of dynamical phases \cite{PRXAlex}. On the other hand, for fundamental oscillations corresponding to the $\beta$-band, $\lambda$ of -0.15$\pi$ and 0.01$\pi$ are obtained for samples S2 and S6 respectively, while the values for the $\gamma$ band are -0.24$\pi$ and -0.17$\pi$. These results suggest that both these bands have near zero Berry phase, where the influence of dynamical phase factors leads to a small deviation from zero.

A Landau fan diagram was also utilized to extrapolate the dominant oscillation phase factor (Fig.~3), where Landau indices were assigned to the positions of the SdH oscillations which correspond to the deepest valleys, for both S2 and S6. The uncertainty of the valley positions is small which doesn't significantly affect the results, and the periodicity of these valleys corresponds to that of the $ \alpha $-band. From linearly fitting the Landau index $n$ as a function of $1/B$, the extrapolated residual Landau indices are $n_0=0.53(5)$ for sample S2 and $n_0=0.52(3)$ for sample S6. These agree well with the above conclusions that the phase factor of the $\alpha$-band, which is the strongest frequency component observed in the quantum oscillations, is close to $\pi$.

Though these results are consistent with those expected for a non-trivial band topology, in Ref.~\cite{PRXAlex} it was suggested that the phase obtained from quantum oscillation experiments cannot directly probe the Berry phase in three-dimensional centrosymmetric materials. In our measurements, the field is applied along the [100] direction, which corresponds to the $ C_4 $ rotation axis of the cubic crystal structure. Due to the symmetry constraints in this situation (mirror symmetry) \cite{PRXAlex}, the spin degeneracy of each band is preserved, and the overall phase factor will be quantized to either $\pi$ or zero, which is not a linear summation of the phase factors of both spin up and down bands, and thus cannot be used to directly obtain the Berry phase or determine the band topology.
Here we note that the same phase factor is found from measurements above and below $ T_ N$. Since the magnetic structure of SmSb has not yet been reported, we are not able to determine if the same symmetry constraints are present in the magnetically ordered state. Electronic structure calculations for the case of zero applied field suggest that SmSb has a trivial band topology but is close to the boundary between trivial and non-trivial \cite{calculation}. However we note that these calculations assume a Sm valence of 3+, while a mixed valence of about 2.75 was reported from x-ray photoemission spectroscopy \cite{campagna1974}. As such it is important to perform further studies to reveal the origin of this nontrivial Berry phase. Evidence for a $\pi$ Berry phase was also found in isostructural LaBi \cite{LaBiIOP}, where both an odd number of band inversions and surface states with an odd number of massless Dirac cones were observed in ARPES measurements \cite{Nayak2017Multiple,LaBiARPES,LaXFDL,XBiARPES}.

In addition, we also performed field-dependent ac susceptibility measurements down to low temperatures where the presence of dHvA oscillations is clearly observed. For comparison, SdH and dHvA oscillations are shown for various temperatures down to 0.3~K in Figs.~4\textbf{a} and 4\textbf{b} respectively, after subtracting the background contribution. The respective FFT are shown in Figs.~4\textbf{c} and 4\textbf{d}, where two principal frequencies are observed, which remain unchanged below $T_N$, similar to previous reports \cite{ozeki1991haas}. The observed values are $F_{\alpha}~\approx~$334~T and 328~T, and $F _{\beta} ~\approx$~633~T and ~$\approx$~590~T for SdH and dHvA respectively, where the differences may be due to a small misalignment. The SdH oscillation amplitudes as $\Delta\rho$ corresponding to both the $\alpha$ and $ \beta $ bands increase significantly with decreasing temperature below 2.5~K, while in dHvA measurements the change is more gradual. Furthermore, the amplitude of the $\alpha$-band oscillations is maximum at around 0.8~K, and upon further decreasing the temperature, the amplitude decreases. Meanwhile, the angular dependence of $F_{\alpha}$ can be fitted with the equation for anisotropic three-dimensional ellipsoidal pockets \cite{3Dellips}(see Supplementary Figure 2), which can describe the anisotropic bullet-like Fermi surface pockets around the X-points, corresponding to the bulk Fermi surface of the $\alpha$-band \cite{calculation,LaBiPRB}.

Figures~5\textbf{a} and 5\textbf{b} present the temperature dependence of the oscillation amplitudes from both SdH and dHvA measurements. Since there is a rapid change of the resistivity both with changing temperature and field, the SdH oscillation amplitudes are displayed as $\Delta\rho$/$\rho$$ _0 $, so as to normalize by the background values. The temperature dependence of the dHvA amplitudes are well fitted across the whole temperature range (0.3 to 10~K) for both the $\alpha$ and $\beta$ bands with the amplitude $\propto R_T$ (Eq.~1), where the fitted values of the effective cyclotron masses are 0.26~m$_e $ and 0.28~m$_e $ respectively. These results are consistent between measurements of both the ac and dc susceptibility (samples S3 and S5 respectively). However, although the SdH and dHvA data coincide well at higher temperatures, below around 2.5~K there is a significant deviation which cannot be accounted for by the LK formula. Moreover, for the $\alpha$-band, the amplitude reaches a maximum at around 1.6~K, before beginning to decrease. We note that these results are highly repeatable, as shown by the measurements of the two samples displayed in Fig.~5, and when the SdH amplitudes are plotted as $\Delta\rho$, the deviation from LK behavior is even more pronounced (see Supplementary Figure 3). In addition, the Dingle temperature analysis does not show a significant change at low temperatures (Supplementary Figure 4), and as a result this behavior cannot be accounted for by the temperature dependence of $T_D$.

One possible explanation for the deviation between dHvA and SdH measurements is that the unusual behavior of SdH oscillations is related to the topologically non-trivial electronic structure which manifests a $\pi$ Berry phase. We note that dHvA measurements of floating zone grown crystals of the proposed Kondo insulator SmB$ _6 $ reveal a steep increase of the oscillation amplitude below 1~K which cannot be explained by LK theory \cite{tan2015unconventional,Hartstein2017}, while there is also evidence for a non-trivial Berry phase \cite{Li1208}. Currently, there is considerable debate over whether the quantum oscillations in SmB$_6$ are from the insulating bulk, or the surface \cite{tan2015unconventional,Li1208}. Proposals such as the magnetic breakdown scenario give rise to quantum oscillations in the bulk of an insulator which deviate from LK behavior \cite{knolle2017excitons}, while it has also been suggested that the anomalous increase of dHvA quantum oscillation amplitudes in SmB$_6$ arises due to surface quantum criticality \cite{erten2016kondo}. In contrast to SmB$_6$, the amplitudes of our dHvA measurements of SmSb can be well fitted using the LK expression and the anomalous behavior is only seen in SdH measurements. This may suggest that the anomalous behavior of SmSb arises only from certain conduction channels with high mobility, which do not significantly contribute to dHvA measurements. Metallic edge states were proposed to explain the anomalous quantum oscillations of some charge transfer salts, where the deviation of SdH amplitudes from conventional behavior is not observed in the dHvA \cite{QHEAnom1,QHEAnom2}. However, these systems have highly conducting quasi-two-dimensional planes where the edge states are suggested to arise in the quantum Hall regime, and this is therefore quite a different scenario to that presented by SmSb. Meanwhile, a comparison between SdH and dHvA cannot be made for SmB$_6$, since SdH oscillations in magnetotransport measurements have not been reported.

Another possibility is that the deviation between SdH and dHvA measurements arises due to the difficulty in disentangling the contributions to the conductivity from the different FS sheets. Although it is often assumed that $\Delta\rho$/$\rho$$ _0 $ is proportional to the density of states and that the temperature dependence is well described by the LK formula, for multiband systems the situation can be more complicated due to the different contributions to the conductivity from each band \cite{shoenberg2009magnetic}. This effect will not influence the dHvA results, which can be well described by the LK formula across the whole temperature range. We note that below around $T_N$, the normalized oscillation amplitude of the $\alpha$-band is anomalously low relative to the dHvA results, while for the $\beta$-band there is an increase. This suggests that with decreasing temperature, there may be changes of the relative contributions to the conductivity from the two bands. Therefore, for systems with multiple FS sheets and large magnetoresistance, analysis of the cyclotron masses using SdH measurements may be difficult. 

Another question posed by these results is why this anomalous temperature dependence of SdH amplitudes is not observed in other $X$Sb compounds. This may be related to the fact that in other $X$Sb materials, the rapid increase of the magnetoresistance occurs at considerably higher temperatures. For example, in LaSb and YSb this occurs at around 80~K and 100~K, respectively \cite{LaSbCava,YSbNew}, while the SdH oscillations are usually measured below 20~K for these materials, where the magnetoresistance shows relatively little change with temperature. However, as displayed in Fig.~1\textbf{b}, the magnetoresistance of SmSb changes strongly below 2~K, which may influence the oscillation amplitudes at low temperatures. Therefore the origin of the dramatic departure from conventional behavior in the SdH amplitudes is an open question, and in particular since this deviation onsets near $T_N$, the possible relationship between the antiferromagnetic state and the anomalous behavior also needs to be explored.

To summarize, we find evidence for a $\pi$ Berry phase in the $\alpha$-band of SmSb from analyzing high field measurements of SdH oscillations. Furthermore, our quantum oscillation measurements show that the amplitudes of dHvA oscillations are well described by the standard LK formula, while those from the SdH effect show anomalous behavior at low temperatures. The origin of this unusual behavior in SdH oscillations remains unclear, and therefore further studies are required to understand the origin of the Berry phase, as well as any relationship with the anomalous SdH oscillations.

\section*{Methods} 

Single crystals of SmSb were synthesized using an Sn flux-method \cite{P1992Growth}. The elements were combined in a molar ratio Sm:Sb:Sn of 1:1:20 in an Al$ _2 $O$ _3 $ crucible before being sealed in evacuated quartz ampoules. These were slowly cooled from 1100$^\circ$C down to 800$^\circ$C, before centrifuging to remove the Sn flux. The resulting crystals are cubic with typical dimensions of around 3~mm. The residual resistivity ratio of $RRR = \rho(300{\rm K})/\rho(0.3{\rm K})\approx4000$, indicates a high sample quality.

Low temperature resistivity and ac susceptibility measurements were performed between 0.27 K and 9 K using an Oxford Instruments $^3$He system, in applied fields up to 15 T. The resistivity data in this system was measured using a Lakeshore 370 resistance bridge, with a current of 0.3 mA across the whole temperature range. Four Pt wires were attached to the sample by spot welding so that the current was applied along the [100] direction. dHvA results were measured using a commercial susceptometer for ac susceptibility measurements, which consists of three sets of coils, including a drive coil, a pick-up coil, and a coil for compensation. Additional resistivity measurements in fields up to 9 T, and dc-susceptibility measurements in fields up to 13 T, were also performed using a Quantum Design Physical Property Measurement System (PPMS) from 300 K to 2 K, where the susceptibility was measured using a vibrating sample magnetometer insert. The high field magnetoresistance measurements were performed up to 60~T using a pulsed field magnet at the Los Alamos National Laboratory, USA, while measurements up to 32~T were carried out using the Cell 5 Water-Cooling Magnet at the High Magnetic Field Laboratory of the Chinese Academy of Sciences.

\section* {Data availability statement}
The data that support the findings of this study are available from the corresponding author upon reasonable request.

\section* {acknowledgments}

We thank Y.~Liu, C.~Cao, Q. Si, Y. Zhou, F. Steglich, P. Li and Z. Z. Wu for interesting discussions and helpful suggestions. This work was supported by the the Science Challenge Project of China (Project No.~TZ2016004), National Key R\&D Program of China (Grants No.~2017YFA0303100 and No.~2016YFA0300202), the National Natural Science Foundation of China (Grants No. U1632275, No. 11604291 and No. 11474251), and the Innovative Program of Development Foundation of Hefei Center for Physical Science and Technology (Project No.~2017FXCX001). A portion of this work was performed at the National High Magnetic Field Laboratory, which is supported by National Science Foundation Cooperative Agreement No. DMR-1644779* and the State of Florida. JS acknowledges support from the DOE BES program "Science of 100 T".

\section*{Competing financial interests} 
The Authors declare no Competing Financial or Non-Financial Interests

\section*{Author contributions} The crystals were grown by F.W.. F.W., C.Y.G., J.L.Z., Y.C. and J.S. performed the measurements, which were analyzed by F.W., C.Y.G., M.S., J.S. and H.Q.Y. The manuscript were written by F.W., C.Y.G., M.S. and H.Q.Y. All authors participated in discussions.

\bibliographystyle{naturemag}

\newpage

\begin{figure}[h]
	\begin{center}
		\includegraphics[width=\columnwidth]{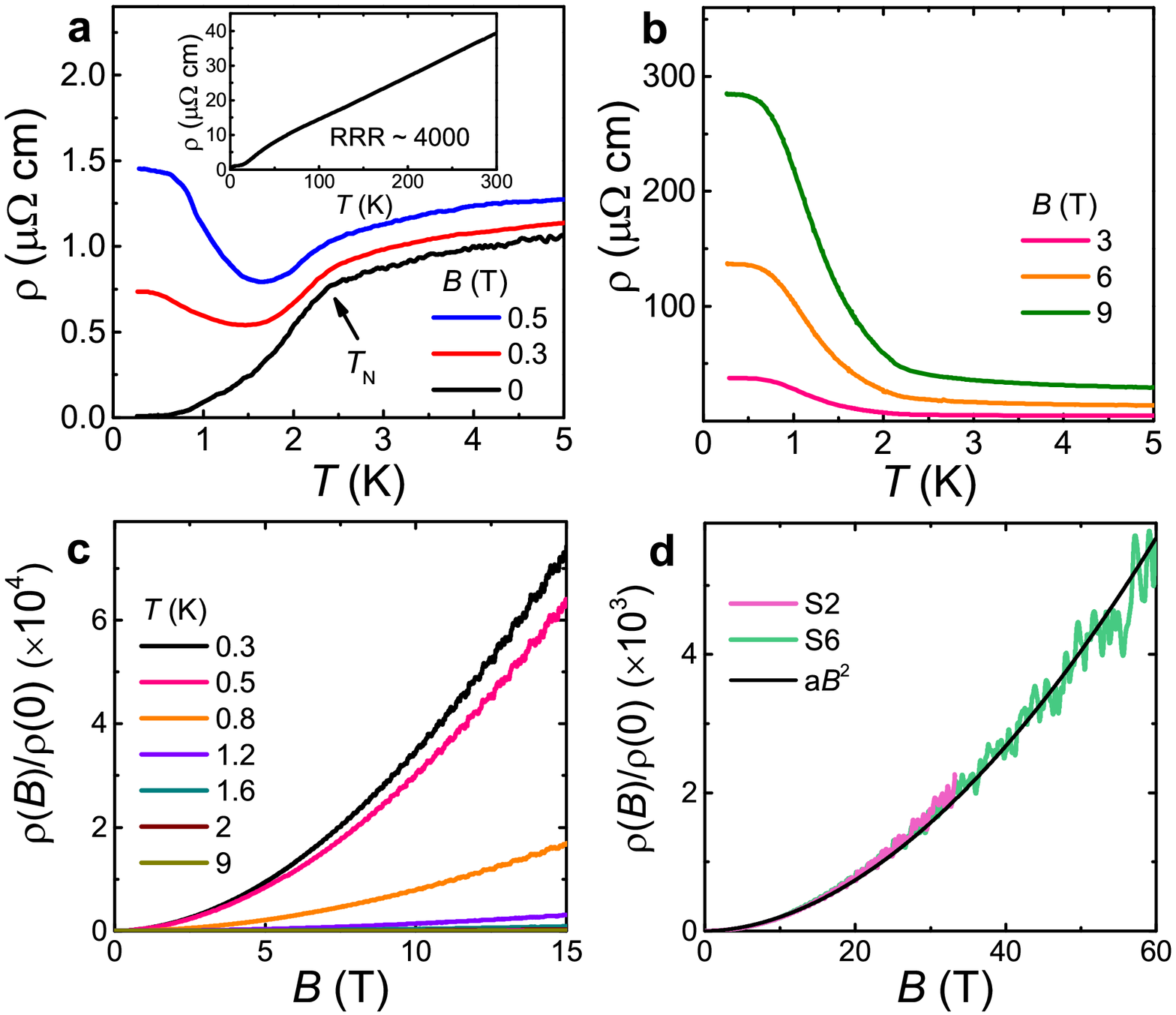}
	\end{center}
	\caption{ \textbf{Resistivity and magnetoresistance of SmSb.} Temperature dependence of the electrical resistivity of SmSb sample S1 at \textbf{a} zero and low applied fields, and \textbf{b,} at high fields. The inset of \textbf{a,} displays the resistivity over the full temperature range from 0.3 to 300~K. \textbf{c,} Field dependence of the transverse magnetoresistance of SmSb sample S1 at various temperatures up to applied fields of 15~T. \textbf{d,} Transverse magnetoresistance at 1.8~K of SmSb samples S2 and S6, up to applied fields of 32~T and 60~T respectively. The solid line shows the $ B^{2}$ dependence.}
	\label{figure1}
\end{figure}

\begin{figure}[tb]
	\begin{center}
		\includegraphics[width=\columnwidth]{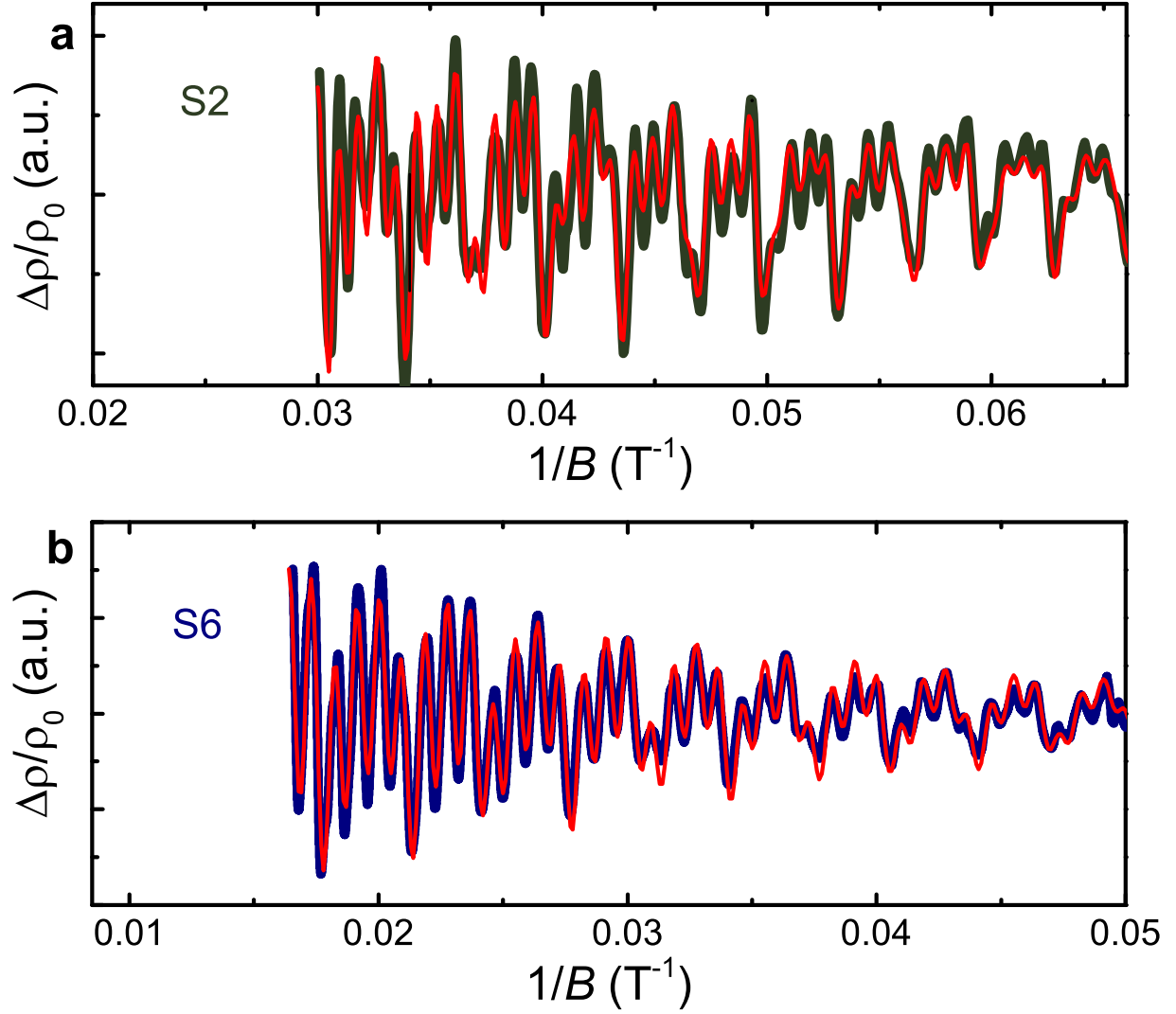}
	\end{center}
	\caption{\textbf{SdH oscillations analysis using Lifshitz-Kosevich theory.} Analysis of the SdH oscillations of SmSb \textbf{a}, samples S2 at 1.8~K and \textbf{b}, S6 at 15~K using Lifshitz-Kosevich theory in Eq.~\ref{equation1}. The red thin line represents the fitted curves and the thick line corresponds to the experimental results. The corresponding field range for fitting is 15-32~T and 20-60~T respectively.}
	\label{figure2}
\end{figure}

\begin{figure*}
	\begin{center}
		\includegraphics[width=1\columnwidth]{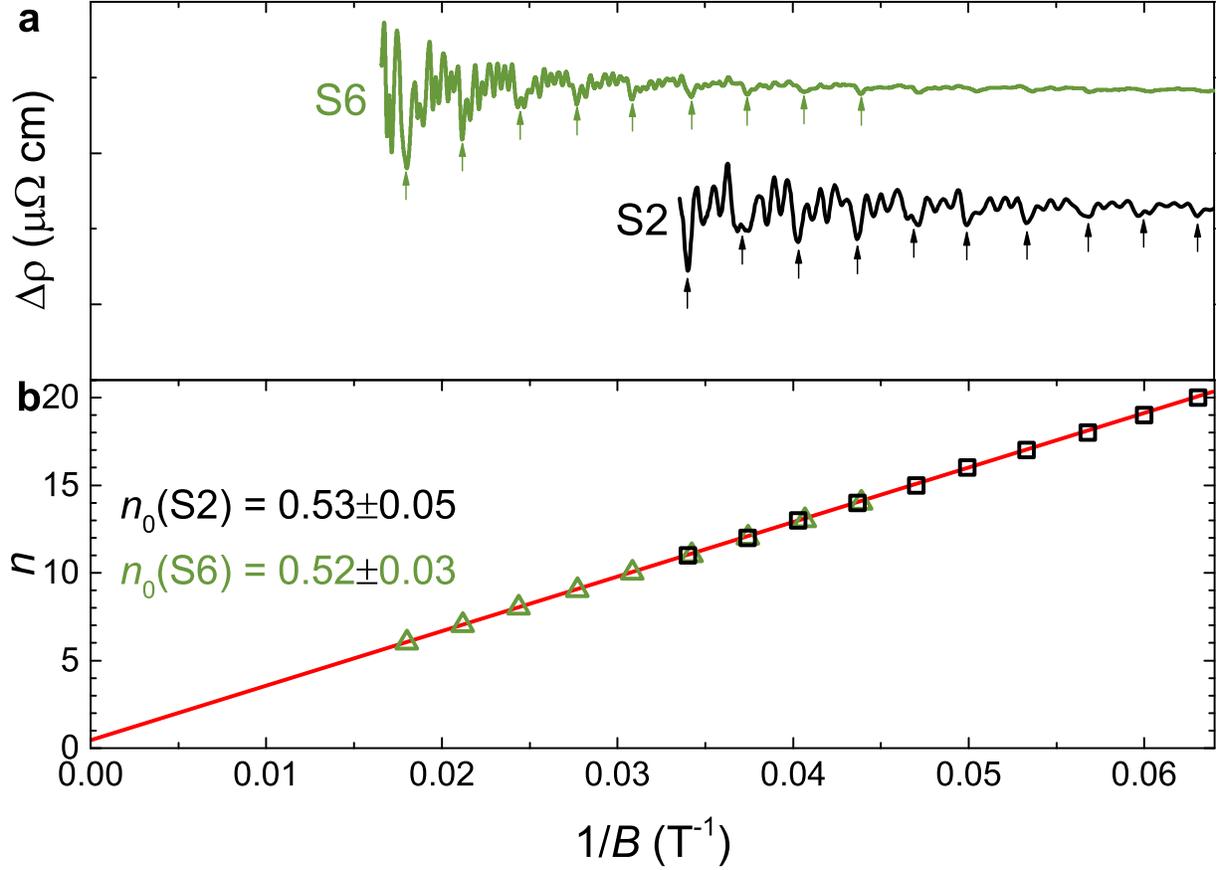}
	\end{center}
	\caption{\textbf{ Landau fan diagram analysis of the Berry phase.
		} \textbf{a,} SdH oscillations of samples S2 and S6 of SmSb plotted as a function of $1/B$. The oscillations were obtained from the field dependence of the magnetoresistance at 1.8~K up to 32~T, and at 4~K up to 60~T, respectively after subtracting the background contribution. \textbf{b,} The $1/B$ dependence of the valley positions in the SdH measurements. The solid lines show linear fits to the data, which both extrapolate to a non-zero residual value close to $n_0=0.5$.}
	\label{figureSX}
\end{figure*}

\begin{figure*}
	\centering
	\includegraphics[width=1\columnwidth]{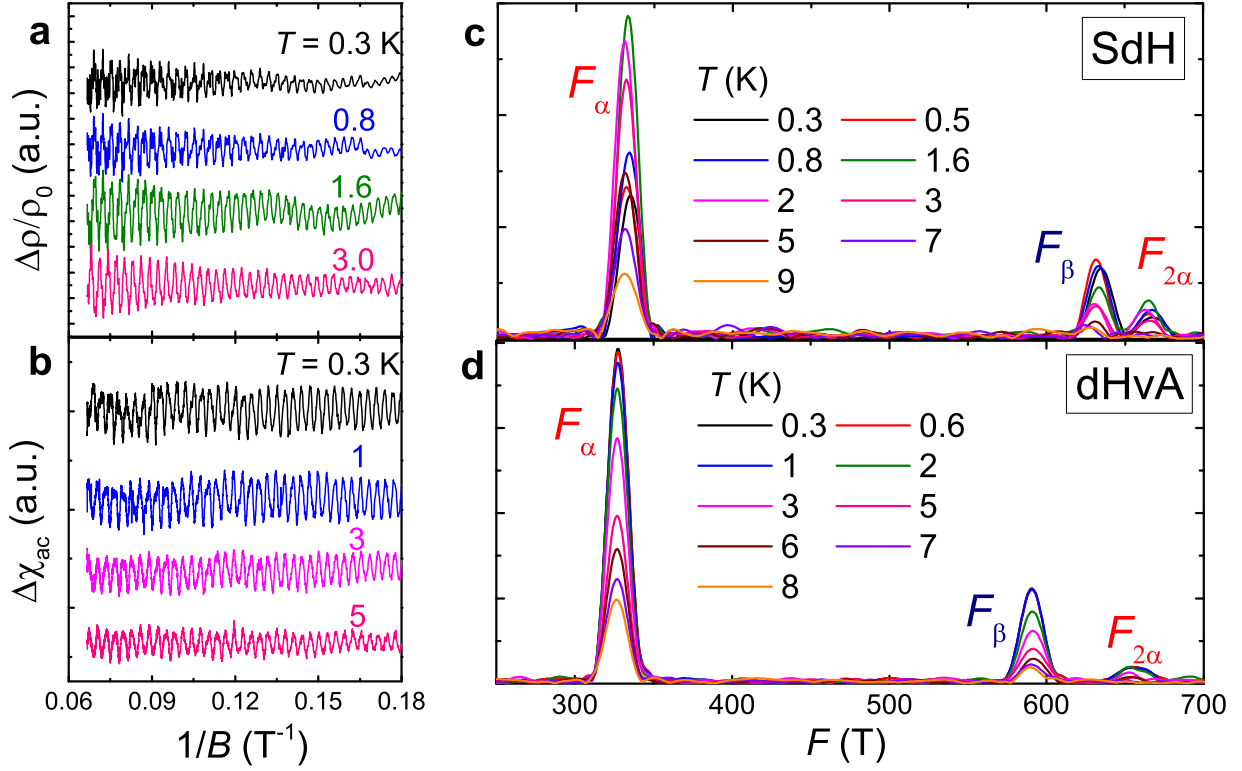}
	\caption{ \textbf{SdH and dHvA oscillations at various temperatures.} \textbf{a,} SdH oscillations obtained from the field dependence of the magnetoresistance of sample S1 of SmSb at various temperatures. \textbf{b,} dHvA oscillations at several different temperatures from field dependent ac~susceptibility measurements of sample S3 of SmSb. The measurements were performed with the magnetic field applied along the principal crystallographic axis. The corresponding FFT spectra are displayed for oscillations in \textbf{c,} SdH, and \textbf{d,} dHvA.}
	\label{figure3}
\end{figure*}

\begin{figure*}
	\begin{center}
		\includegraphics[width=1\columnwidth]{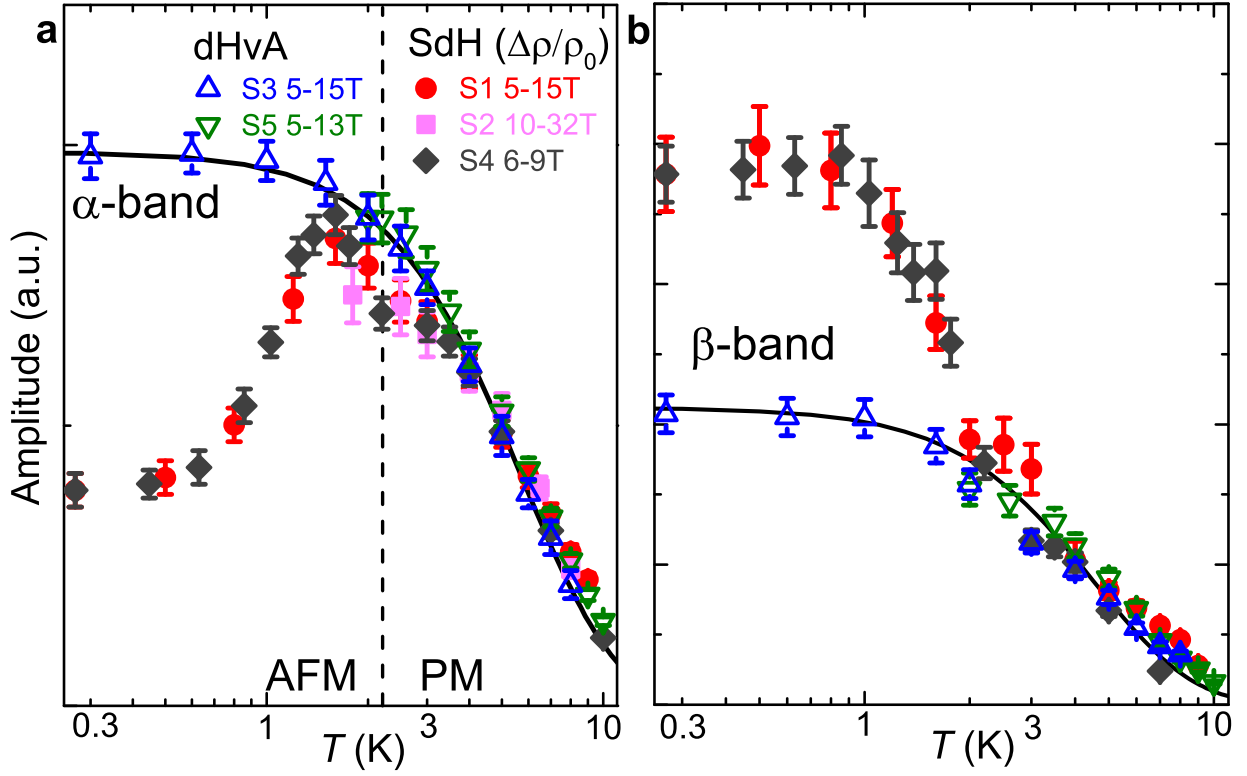}
	\end{center}
	\caption{ \textbf{Anomalous temperature dependence of SdH oscillation amplitudes.} Temperature dependence of the SdH (filled symbols) and dHvA (empty symbols) amplitudes of the fundamental \textbf{a,} $\alpha$-band and \textbf{b,} $\beta$-band frequencies, for several different samples, where the SdH amplitudes are displayed as $\Delta\rho$/$\rho$$ _0 $. Samples S1 and S3 were measured in a $^3$He cryostat in fields up to 15 T, S2 was measured at the High Magnetic Field Laboratory, while S4 and S5 were measured using a Physical Property Measurement System. The field ranges over which the data were analyzed are displayed, while the vertical dashed line marks the position of $T_N$. The solid lines display fits to the dHvA amplitudes using the temperature damping factor of the Lifshitz-Kosevich formula, with effective masses of 0.26~m$_e $ for the $\alpha$-band and 0.28~m$_e $ for the $\beta$-band.}
	\label{figure4}
\end{figure*}

\end{document}